\documentclass[english,latin9]{aa}
\usepackage[T1]{fontenc}
\usepackage{graphicx}

\makeatletter

\newcommand{\noun}[1]{\textsc{#1}}

\titlerunning{The potential of asteroseismology for probing the core of
  white dwarfs}
\authorrunning{Giammichele et al.}

\makeatother

\usepackage{babel}
\begin{document}


\title{The potential of asteroseismology for probing the core chemical
  stratification in white dwarf stars}

\author{N. Giammichele\inst{1,2} \and S. Charpinet\inst{1,2} \and  P. Brassard
\inst{3}\and  G. Fontaine\inst{3}}

\institute{Université de Toulouse, UPS-OMP, IRAP, Toulouse, F-31400,
  France\and CNRS, IRAP, 14 avenue Edouard Belin, F-31400 Toulouse, France\\
\email{{[}noemi.giammichele,stephane.charpinet{]}@irap.omp.eu}\and Département
de Physique, Université de Montréal, CP 6128, Succursale Centre-Ville,
Montréal, QC H3C 3J7, Canada\\
\email{{[}brassard,fontaine{]}@astro.umontreal.ca}}

\offprints{N. Giammichele}

\date{Received 21 October 2016; Accepted 8 November 2016}

\abstract{The details of the C/O core structure in white dwarf stars has
  mostly remained inaccessible to the technique of asteroseismology,
  despite several attempts carried out in the past.}{We re-assess the
  potential of asteroseismology for probing the chemical stratification
  in white dwarf cores, in light of new highly efficient tools recently
  developed for that purpose.}{Using the forward modeling approach and
  a new parameterization for the core chemical stratification in ZZ Ceti
  stars, we test several situations typical of the usually limited
  constraints available, such as small numbers of observed independent
  modes, to carry out asteroseismology of these stars.}{We find that,
  even with a limited number of modes, the core chemical stratification
  (in particular, the location of the steep chemical transitions expected
  in the oxygen profile) can be determined quite precisely due to the
  significant sensitivity of some confined modes to partial reflexion
  (trapping) effects. These effects are similar to the well known
  trapping induced by the shallower chemical transitions at the edge of
  the core and at the bottom of the H-rich envelope. We also find that
  success to unravel the core structure depends on the information
  content of the available seismic data. In some cases, it may not be
  possible to isolate a unique, well-defined seismic solution and the
  problem remains degenerate.}{Our results establish that constraining
  the core chemical stratification in white dwarf stars based solely on
  asteroseismology is possible, an opportunity that we have started to
  exploit.}


\maketitle

\section{Introduction}

For more than two decades, the main outcome of quantitative asteroseismic
studies of white dwarf stars has been to derive the fundamental parameters
(mass or surface gravity, effective temperature) and the most superficial
chemical structure (the H-rich and/or He-rich layer thickness) of these
stars (see, e.g., Section 8 of \citealt{fb08} {and other reviews on the
subject from \citealt{2008ARA&A..46..157W} and \citealt{2010A&ARv..18..471A}}).
The deeper structures,
in particular the core chemical stratification, have been considered
either firstly, as secondary due to the perceived low sensitivity of
$g$-modes to
these deep regions, or secondly, as features that need to be set a priori,
using most notably the results of full evolutionary calculations, in
order to perhaps achieve more accurate determinations of the parameters
mentionned above \citep[see, e.g.,
][]{2002ApJ...581L..33F,2005MNRAS.363L..86M,2008MNRAS.385..430C,2009MNRAS.396.1709C,2011MNRAS.414..404B,2012MNRAS.420.1462R,2014ApJ...794...39B,2016ApJS..223...10G}.

{However, \citet{2002A&A...387..531C} and later on \citet{2005A&A...439L..31C}
showed from detailed studies of mode trapping in two representative
white dwarf and pre-white dwarf evolutionary models that some low-order,
low-degree $g$-modes can be very sensitive to the detailed structure of the
chemical stratification in the white dwarf core.
Moreover}, recently, \citet{2015ApJ...815...56G,2016ApJS..223...10G},
in a thorough re-analysis of the two pulsating ZZ Ceti stars and
spectroscopic twins GD~165
and Ross~548, quite unexpectedly found that the pulsations detected
in GD~165 have only a weak dependence on the bulk composition of the
core, while those observed in Ross~548 are quite sensitive to a variation
of this quantity. This puzzle found a natural explanation in that all the
modes in the derived seismic model for GD~165 have amplitudes and
weight functions that do not extend into the deep core, while three
of the six modes associated with the periods observed in Ross~548 are
partly confined in the core, and therefore bear a strong sensitivity
on its composition. {These discoveries open up} the most interesting
possibility that details of the core composition stratification in white dwarf
stars could also be probed with asteroseismology using such deeply
confined modes. It is with this possibility in mind that we started
developing the new numerical tools described at length in
\citet{2016ApJS..inpress}.

\begin{figure*}
\begin{centering}
\includegraphics[scale=0.55]{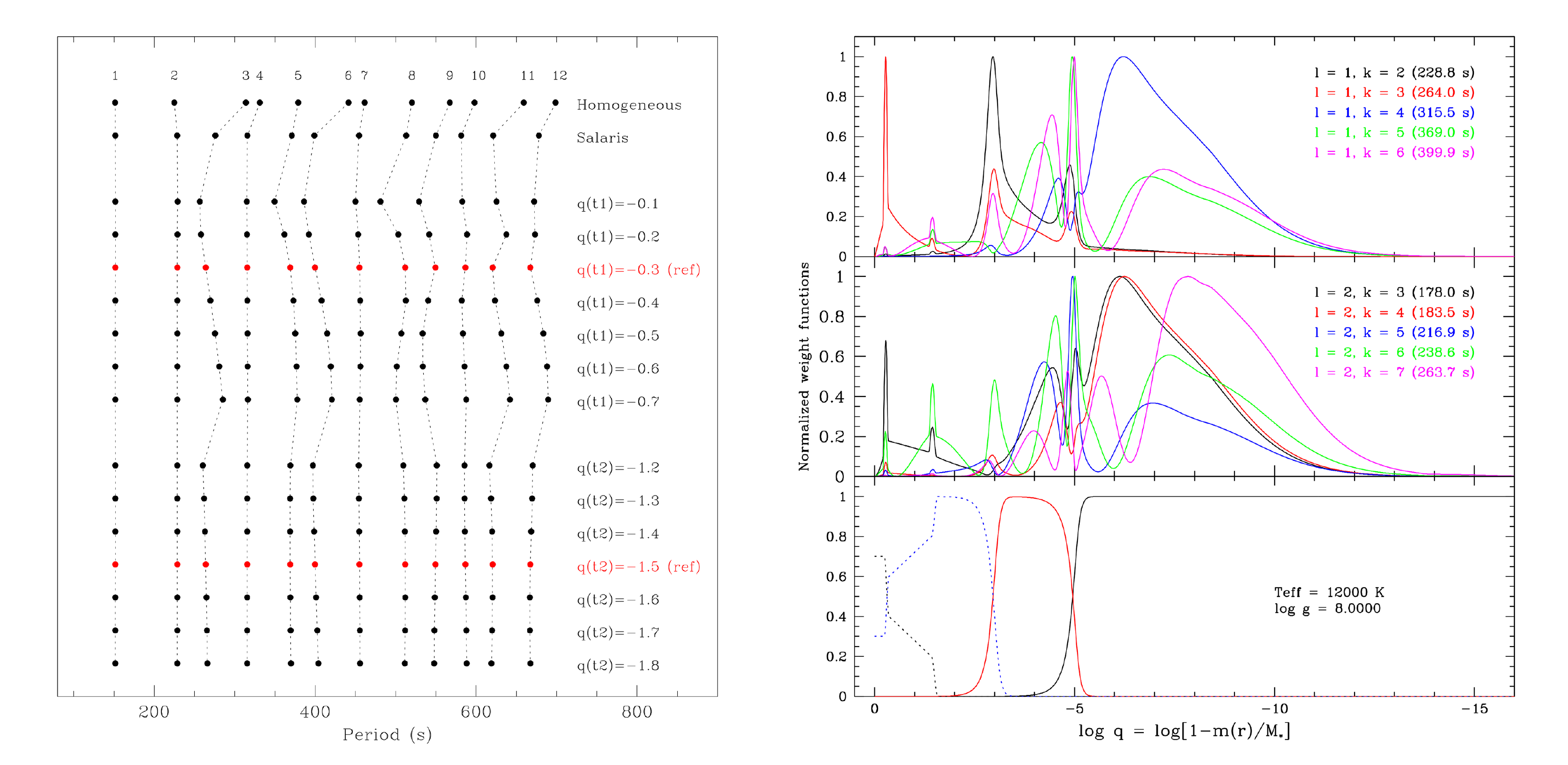}
\par\end{centering}

\caption{\emph{Left panel} : Period spectra of the $\ell=1$, $k=1-12$ $g$-modes
for a series of models with different core configurations (see text
for details). Modes of same radial order are connected by a dotted
line to ease following period changes. \emph{Right panel }: Oxygen
(black dotted line), carbon (blue dotted line), helium (red solid
line), and hydrogen (black solid line) stratification in the reference
model (bottom subpanel). The top and middle subpanels show the weight
functions for a subset of dipole ($\ell=1$) and quadrupole ($\ell=2$)
$g$-modes computed for this reference model. \label{fig1}}
\end{figure*}

Based on such tools, we assess, in this paper, the potential of
asteroseismology for probing {quantitatively} the core chemical
stratification in white dwarf stars,
in particular for situations typically encountered in the field when
only a relatively small number of modes is available for a seismic
analysis. We illustrate in Section 2 the typical signature of the
core stratification on the oscillations periods and discuss in Section
3 a series of tests to evaluate whether this information contained
in the period spectrum can be recovered with our approach to white
dwarf asteroseismology or not. We conclude in Section 4.

\section{Seismic signature of the core stratification}

In \citet{2016ApJS..inpress}, we introduced a new generation of hydrostatic
equilibium white dwarf models designed specifically for in-depth asteroseismic
probing of the stellar structure. These new models most notably provide
a new parameterization for the C/O stratification allowing us to
explore a vast range of chemical configurations in the core. In particular,
they can closely mimic the double-ramp-like oxygen profiles predicted
by standard evolutionary calculations, including cases with a triple
transition zone at the core edge (see the details provided in
\citealp{2016ApJS..inpress}). Here, we take advantage of this new tool
to {characterize} the sensitivity of low-degree, low-order $g$-modes (most
often observed in white dwarf pulsators) to the core C/O
stratification. We isolate, in particular, the effects induced by abrupt
changes in the core composition.

The left panel of Fig.~\ref{fig1} shows the $g$-mode period spectrum
(concentrating on low-order dipole, $\ell=1$, modes) for a series
of similar DA white dwarf models, except for their C/O core structure.
The first two models of the series have, respectively, a homogeneous
C/O core (75\% oxygen in mass) and a stratified oxygen profile obtained
from the evolutionary calculations of \citet{2010ApJ...716.1241S}. Important
differences in the distribution of the $g$-mode periods between these
two models clearly emerge, showing that the impact of the core structure
can indeed be significant, {an observation in line with the conclusions
of \citet{2002A&A...387..531C} and \citet{2005A&A...439L..31C}}.

\begin{figure*}
\begin{centering}
\includegraphics[scale=0.85]{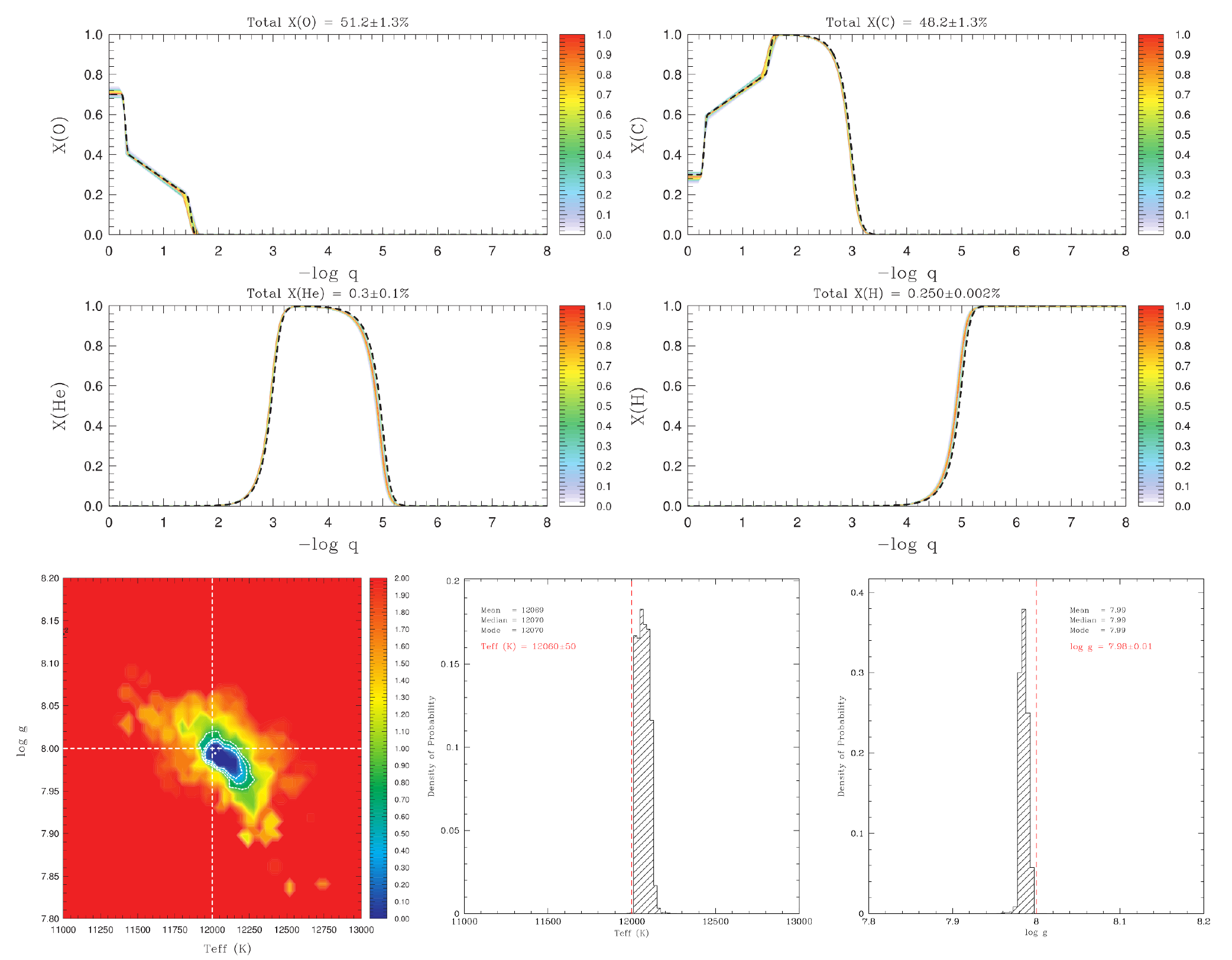}
\par\end{centering}

\caption{Seismic solution obtained for our artificial star using ten periods.
From top to bottom, left to right, panels show : 1) the derived probability
distribution functions normalized for the oxygen, carbon, helium,
and hydrogen profiles with, in each case, the true profile from the
reference model superimposed as the black dashed curve, 2) the projected
$S^{2}$-map (in log scale) for the two global parameters $T_{{\rm eff}}$
and $\log g$ with the true values indicated by the white vertical
and horizontal dotted lines, and 3) the probability distribution functions
derived for these two parameters, with the vertical red dashed line
indicating the true value (see text and \citealt{2016ApJS..223...10G}
for further details).\label{fig2}}
\end{figure*}

With the following models, using the new core parameterization from
\citet{2016ApJS..inpress}, we identify the main effects perturbing
the eigenmode periods. The right panel of Fig.~\ref{fig1} (bottom subpanel)
shows the distribution of oxygen (as well as carbon, helium, and hydrogen)
in one such models, that we now on refer to as our reference structure.
It illustrates the typical double-ramp shape of the oxygen mass fraction
characterized by two steep descents. The next seven models represented
in the left panel of Fig.~\ref{fig1} reveal how the $g$-mode period
spectrum reacts to a change in the position, $q(t_{1})$ (also named
``$t_{1}$\_lq'' in \citealt{2016ApJS..inpress}), of the first
drop in the oxygen profile going outward. We find that some modes (e.g.,
the $k=2$
and $k=4$ modes) remain clearly unaffected, while others (e.g., the
$k=3$ mode) experience significant changes of their periods. This
is understood by looking at the structure of the modes through, for example,
their so-called weight functions \citep[see][]{1992ApJS...80..369B}
represented, for our reference model, in the right panel of Fig.~\ref{fig1}
(top and middle subpanels). This quantity identifies the regions inside
the star that contribute to the period of the mode, thus indicating which
part it is sensitive to. In particular, we note that some of the modes
have a significant peak arising exactly at the position of the first
drop in the oxygen profile (e.g, the $\ell=1$, $k=3$ mode) while
others show no amplitude at all there (e.g, the $\ell=1$, $k=2$
and $k=4$ modes). The latter (or former) are indeed the modes showing
no (or a strong) sensitivity to the parameter specifying the position of
this transition.
We point out that very similarly, though more or less pronounced,
peaked structures associated with the chemical transitions between
C/O and He, and between He and H, are also seen in these weight functions,
thus indicating a specific sensitivity of the modes to the core-mantle
and mantle-envelope boundaries. This well known sensitivity has indeed
already permitted to measure the hydrogen and helium layers thickness
in pulsating white dwarfs using asteroseismology. It is due to partial
wave reflections at the chemical interfaces causing modes to be either
trapped in the upper layers (e.g., in the envelope) or confined below
the transitions (e.g., in the core). These trapping and confinement effects,
including their impact on the period distribution, have been studied
in detail, e.g., by \citet{1992ApJS...80..369B}. Therefore, we find
here that an additional trapping phenomenon is at work due to the
deeper chemical gradients expected in the core, which primarily affects
modes that are already confined deeper in the star by the other shallower
chemical transitions.
{This trapping effect occurring in white dwarf cores was formerly
identified by \citet{2002A&A...387..531C} who reported
similar structures in the weight functions of some $g$-modes computed
from a representative fully evolutionary DA white dwarf model. These authors demonstrated, with numerical experiments inspired from
\citet{1992ApJS...81..747B} and \citet{2000ApJS..131..223C}, that this effect
could have significant impact on the period spectrum.}
Consequently, there is also a specific signature
of the deep core chemical stratification in the period spectrum, which,
at least in principle, should be accessible to asteroseismology.
{This may for instance allow us to place constraints on
physical processes such as core overshooting during pre-white dwarf stages,
as suggested by \citet{2005A&A...439L..31C}}.

The last series of seven models shown in the left panel of Fig.~\ref{fig1}
illustrates the impact of changing the position, $q(t_{2})$ (also named
``$t_{2}$\_lq'' in \citealt{2016ApJS..inpress}), of the second
drop in the oxygen profile. The effect on the period spectrum is obviously
less pronounced in this case but a close comparison still reveals
some noticeable variations. This lower sensitivity associated to this
particular transition is also revealed by the weight functions of
the modes (right panel of Fig.~\ref{fig1}) that usually only show
a small secondary peak in that region. However and most importantly,
the variations induced on the g-mode frequencies within this series
of models can still reach up to 80 $\mu$Hz (e.g., for the $\ell=1$,
$k=3$ mode whose period varies from 260.5 s to 266 s), which is significantly
larger than the precision obtained on measured frequencies, even
from relatively short ground-based campaigns. The subtle information
relative to this transition is therefore clearly present and, in principle,
also accessible to asteroseismic probing.

\begin{figure*}
\begin{centering}
\includegraphics[scale=0.85]{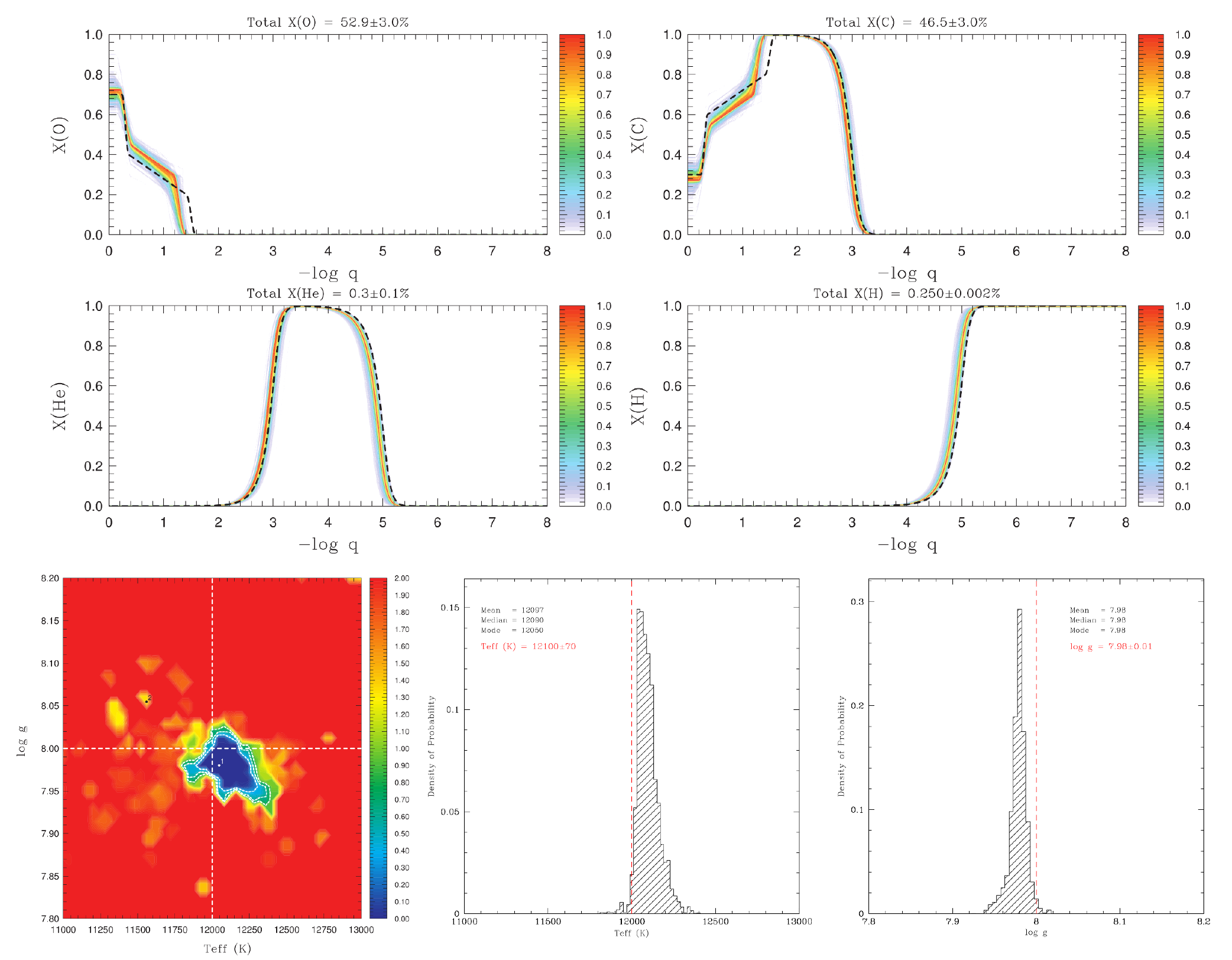}
\par\end{centering}

\caption{As Fig. \ref{fig2}, but with a set of five periods that leads to
a well-defined solution.\label{fig3}}
\end{figure*}

We end this section by pointing out that because some of the modes,
through trapping effects, have a significant sensitivy to the internal
structure of the core, models that assume a given stratification,
e.g., those obtained from detailed evolutionary calculations, make a rather
strong assumption
affecting the calculated period spectrum. Even a relatively small
shift in the positions of the two oxygen drops would result in period
variations that are significant, exceeding largely the precision at
which the periods are measured. It can be problematic when such models
are used for precise asteroseismic inferences. Unless the real core
stratification is very accurately known (though the many uncertainties
associated with evolutionary calculations makes this very unlikely), it could
potentially bias solutions, or at best limit the accuracy at which
the models can match the periods and constrain the other structural
parameters of the star.

\section{The potential of asteroseismology}

The specific dependency of the $g$-mode period spectrum to the core
chemical stratification discussed above opens up the highly interesting
opportunity to constrain the core structure in pulsating white dwarf
stars directly from the observed oscillation modes, independently from
any evolutionary calculations. To estimate the potential of asteroseismology
for that purpose, we propose a series of three tests. We consider
our reference model, with parameters and chemical stratification given
in the right panel of Fig.~\ref{fig1} (bottom subpanel), as an ``artificial
star'' and compute the pulsation periods associated with this structure.
We then apply to each frequency a random perturbation following an
unbiased normal distribution with $\sigma=0.1$ $\mu$Hz that mimics
typical measurement errors from ground based data. Subsets of the
perturbed periods are then selected to form the sets of ``observed
periods'' now considered as if they were typical observations and
submitted to our asteroseismic procedure. The latter is described
in detail in \citet[and reference therein]{2016ApJS..223...10G}.
In a nutshell, it is a double optimization procedure using the forward
modeling approach which is solved with an efficient massively parallel
multimodal optimization code (named \noun{Lucy}) based on an hybrid
genetic algorithm. This tool allows us to recover efficiently the
global solution of the optimization problem, but also any secondary
solutions that may occur if the problem turns out to be degenerate.
This will be essential as shown below. This code can also deal with
high dimension (many parameters) problems, in particular since it
does not rely on grid computations (see, e.g., \citealt{2015ASPC..493..151C}),
which also is crucial in the present context.

\begin{figure*}
\begin{centering}
\includegraphics[scale=0.85]{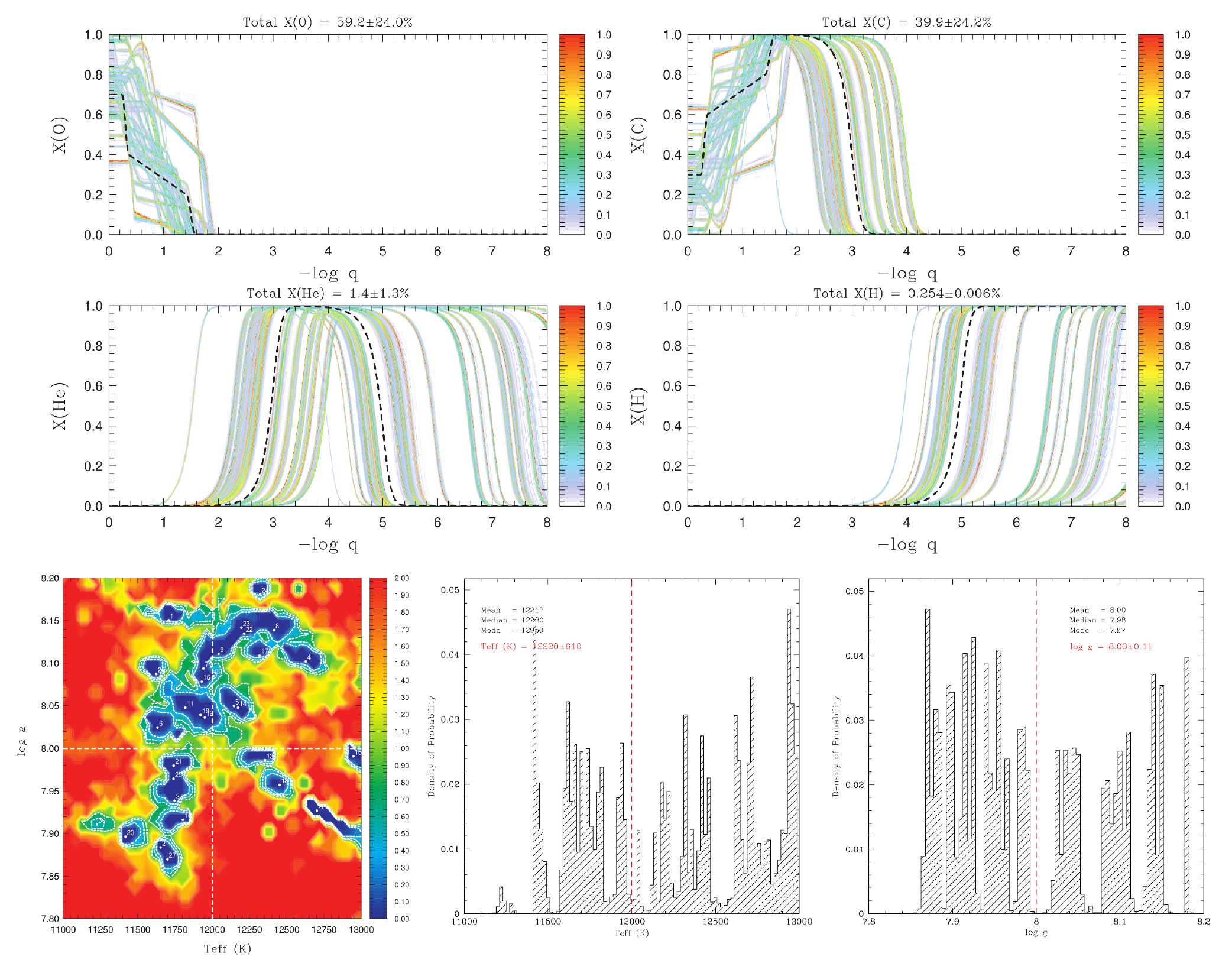}
\par\end{centering}

\caption{Same as Fig. \ref{fig2}, but with a set of five periods that leads to
degenerate solutions.\label{fig4}}
\end{figure*}

The analyses are conducted using the new core parameterization described
and tested in \citet{2016ApJS..inpress}. Along with the global parameters
$T_{{\rm eff}}$ and $\log g$, and the parameters defining the helium
and hydrogen distribution, the optimization problem consists of best-matching
the set of observed periods with up to 13 parameters to adjust.
We stress that most of these parameters are shape parameters allowing
to explore with enough flexibility various composition profiles, as
explained in \citet{2016ApJS..inpress}. The main question arising
with this setup is whether the problem is sufficiently constrained
to allow for a unique determination of the chemical composition in
the core (and elsewhere), or if, on the contrary, the problem remains
undetermined with strongly degenerate solutions. The answer certainly
depends on the number of observed periods available to constrain the
model, as well as the information content of these periods (low-order
modes being generally the most significant from that point of view). In
the following, we explore three configurations involving, for
the first one, a set of ten observed periods, and different sets of
five observed periods for the last two experiments. These numbers are
not chosen arbitrarily and cover the typical situations that one has
to deal with for asteroseismic studies of pulsating DA white dwarfs,
since most of these pulsators, especially those close to the blue
edge, typically show between five to ten modes only.

The first test is done with ten observed periods, namely the modes
$\ell=1,$ $k=2,3,4,5,6$ and $\ell=2$, $k=3,4,5,6,7$ (all shown
in the right panel of Fig. \ref{fig1}). The search for a seismic solution
was performed, as usual, on a very wide range of model parameters
covering amply all possible pulsating DA white dwarf structures and
assuming that only modes of degree $\ell=1$ and 2 are visible. Besides
this, no a priori assumption is made on the identification of each
mode, which is therefore free during the optimization. The result
of the search is summarized in Fig.~\ref{fig2}. We find that both
the global parameters and the composition stratification of oxygen,
carbon, helium, and hydrogen from the original model constituting
our artificial star are well recovered at a rather high level of internal
precision, according to the probablity distribution functions.
We point out here that the derived profiles shown in
the top four panels of Fig.~\ref{fig2} are also probability distribution
functions of the plotted quantity (normalized to one at maximum)
at each $\log q$ position. These
are built from the global sampling and evaluation (in terms of best
matching the periods) of many different stratifications (virtually
several hundred of thousands of models in this specific case) made
by the optimization code during the search for the optimal solution.
We also find that the periods are well recovered, largely within the
errors induced by the random perturbations applied to the original
set of frequencies. This first test is therefore extremely encouraging,
since a well-defined unique solution is found and recovers correctly
the stratification of the star, in particular in the core.

The second test explores the effect of using only five modes for the
analysis. We chose the modes $\ell=1$, $k=2,4,10$ and $\ell=2,$
$k=7,8$ for that purpose. Running the optimization procedure in the
exact same manner as before, we find the results summarized in Fig.~\ref{fig3}.
Again, a unique solution clearly emerges in this analysis,
although with an increased spread of the probability distribution
functions (larger internal errors). Overall, the original model structure
and parameters are very well recovered. We note a small shift particularly
in the determination of the position of the second drop in the oxygen
profile, which we attribute to a small bias introduced when perturbing
the periods randomly. Indeed, even if these perturbations are generated
as unbiased normal deviates, we consider here only five periods which
represents rather small-number statistics. Moreover, as discussed in the
previous section, the sensitivity of the modes to this particular
chemical transition is weaker, meaning that a small bias in the period
spectrum could more noticeably affect its determination from asteroseismology.
The periods are also well matched in that case. This test shows that
even with a reduced number of modes, a well-defined seismic solution
can be found, allowing us to recover the chemical stratification of the star
without ambiguity (but with a lower overall precision). The problem
turns out to be still sufficiently constrained in that case.

However, the third and last test of the series underlines that it
might not always be the case. This experiment considers another set
of five observed periods, namely the modes $\ell=1$, $k=3,6,8$ and
$\ell=2$, $k=3,6$, which leads to the results illustrated in Fig.~\ref{fig4}.
The well-defined solution of the former test has now
disappeared to leave instead a strongly degenerate set of equivalent
solutions, all matching at the highest possible precision the observed periods.
This case demonstrates, if need be, that the multimodal optimization code
indeed properly finds the multiple solutions when the problem is degenerate.
The probability distribution functions are no longer well localized and obviously
nothing can be recovered about the original structure of the star.
Hence, in this situation the problem is clearly underdetermined,
albeit using the same number of modes as the previous test shown
in Fig.~\ref{fig3}. We point out that some of these degeneracies
could probably be lifted, but at the expense of additional
constraints or assumptions on the model structure (such as assuming a
core oxygen profile, for example).

{Finally, we point out that we do not encounter ambiguities in our
tests that could be related to an inherent core-envelope symmetry discussed
in \citet{2003MNRAS.344..657M}.
These authors suggest that a structural degeneracy of the seismic solution
may exist due to the way pulsation modes sample the core and the envelope of the star. We clearly do not have this degeneracy in our two well constrained cases where the original model is accurately and unambiguously recovered. We see at least two reasons explaining why this core-envelope symmetry does not manifest itself here. First, the \citet{2003MNRAS.344..657M} analysis applies to high overtone
modes which are in the asymptotic regime, while in our tests, modes of low to mid
radial order, typical of ZZ Ceti stars close to the blue edge of the instability strip, are considered. Lower overtone modes are much less affected by this symmetry. Second and more importantly, the core-envelope symmetry is approximate in realistic stellar models, as discussed by the authors themselves.
This means that the period distribution in the "symmetrical model"
does not exactly match the period distribution of the true stellar structure.
In a seismic analysis context, this translates into different
values for the merit function quantifying the quality of the fit. Consequently,
if a sufficiently accurate period fit can be achieved, the solution corresponding
to the true structure would dominate in terms of a significantly better value of
the merit function, thus lifting any ambiguity. In the case of the DB pulsator GD358 discussed in \citet{2003MNRAS.344..657M}, the two incompatible best fit solutions proposed by \citet{2003ApJ...587L..43M} and \citet{2002ApJ...581L..33F} have
average rms period residuals of $\sigma_P \sim 1$s which, especially in light of some of the tests carried out in \citet{2016ApJS..inpress}, may simply be
insufficient to clearly separate the structural
signatures of the core and of the envelope. The quality of fit
achieved in our present tests and in applications of our approach
to real white dwarf pulsators (see, e.g, \citealt{2015IAUGA..2256942G} and
forthcoming papers in preparation) is much better and strongly alleviates the risk
of such a degeneracy. This underlines the fact that obtaining high fit accuracy,
ideally down to the precision of the measured periods, is essential to exploit the potential of probing the core stratification in white
dwarf stars from asteroseismology.
}

\section{Conclusion}

Our examination of the specific sensitivity of low-order, low-degree
$g$-modes to the chemical stratification in the core of white dwarf
stars and the series of ``Hare \& Hounds'' tests allow us to draw
two important conclusions about the potential of asteroseismology
for deciphering the core properties of these stars. First, we clearly
find that the sensitivity of the modes to these deep regions can be
exploited to unravel their chemical stratification, even with relatively
few modes as often observed for pulsating white dwarfs. However, success
or failure to do so does not really depend on the actual number of
available modes (although having more modes certainly help), but more
on which specific modes are observed and used for the seismic analysis. As
demonstrated by our tests, the analysis of a white dwarf pulsator
showing only 5 modes
could as well result in a single well-defined solution unravelling
most of its structure or in a useless set of highly degenerate solutions
leaving no real clue on the actual structure of the star. This behavior
is certainly linked to the individual properties of the modes (illustrated
in particular by their weight functions) and the ``differential''
information contained in the distribution of these modes relative
to each other. However, predicting when a set of observed periods
leads to strong constraints on the white dwarf core structure remains
difficult. Each star should be considered as a specific case and it
is only after performing the detailed seismic analysis that it shall
be possible to state whether the problem is sufficiently constrained
or underdetermined.

We also point out that our tests very clearly refute a statement often
heard in the community that models defined with more parameters than
the number of observed periods cannot be constrained (i.e., are
underdetermined). This widespread misconception of statistics is nicely
discussed, e.g., by statistician and Pr. J. Vanderplas in his ``Model
Complexity Myth''
article\footnote{https://jakevdp.github.io/blog/2015/07/06/model-complexity-myth/}.
We cannot refrain but quote the very first sentences of that article:
``An oft-repeated rule of thumb in any sort of statistical model fitting
is \textquotedbl{}you can't fit a model with more parameters than
data points\textquotedbl{}. This idea appears to be as widespread
as it is incorrect. On the contrary, if you construct your models
carefully, you can fit models with more parameters than data points.''
The faulty assertion in fact holds true only for simple linear
systems, such as the well known problem of fitting a set of measurements
(of any kind) with polynomials of the type $y(x)$. Note that ``linear''
here refers to the linearity of model parameters (e.g., the coefficients
of a polynomial expression) rather than linearity of the dependence on
the data $x$. In a seismic context, the response of the pulsation
periods on the variation of one of the defining parameters is, in fact,
highly nonlinear and mode-dependent. Hence, the complex optimization
problems that we encounter in asteroseismology involving pulsation
computations on detailed stellar structures are clearly not linear
and do not fit into this class of simple problems.

Observationally speaking, asteroseismology of pulsating white dwarfs
is currently under a revolution with very high quality data delivered
from space by the \emph{Kepler} satellite (original mission) and the
still operating Kepler 2 (K2) mission. We also envision that the \noun{Tess}
project will further feed the field with ultra high precision seismic
data for many more white dwarfs in a close future. In this context,
the potential of asteroseismology for probing the core stratification
in white dwarf stars will soon come to life. Detailed analyses
of specific objects to achieve this goal {have already been briefly
exposed, for example, in \citet{2015IAUGA..2256942G}, and will be fully presented
in a series of forthcoming articles}.

\begin{acknowledgements}
S. Charpinet acknowledges financial support from \textquotedblleft Programme
National de Physique Stellaire\textquotedblright{} (PNPS) of CNRS/INSU,
France, and from the Centre National d\textquoteright Études Spatiales
(CNES, France). This work was granted access to the HPC resources
of CALMIP under allocation number 2016-p0205. This work was supported
by the FQRNT (Québec) through a postdoctoral fellowship awarded to
N. Giammichele. G. Fontaine also acknowledges the contribution of
the Canada Research Chair Program.
\end{acknowledgements}

\end{document}